\documentclass[12pt]{iopart}
\usepackage{bm}
\usepackage{color}
\usepackage{amssymb}
\usepackage{graphicx}
\usepackage{esint}

\begin{document}

\title{Solitary structures with ion and electron thermal anisotropy}

\author{Murchana Khusroo and Madhurjya P Bora}

\address{Physics Department, Gauhati University, Guwahati
781014, India.}
\ead{mpbora@gauhati.ac.in}
\vspace{10pt}
\begin{indented}
\item[]April 2015
\end{indented}

\begin{abstract}
Formation of electrostatic solitary structures are analysed for a magnetised plasma with ion and electron thermal anisotropies. The ion thermal anisotropy is modelled with the help of  the Chew-Goldberger-Low (CGL) double adiabatic equations of state  while the electrons are treated as inertia-less species with an anisotropic bi-Maxwellian velocity distribution function. A negative electron thermal anisotropy $(T_{e\perp}/T_e{\parallel}>1)$ is found to help form large amplitude solitary structures which are in agreement with observational data.
\end{abstract}

%
%
%
%
%

\section{Introduction}

Solitary structures, commonly known as electrostatic solitary waves
(ESWs) or Solitons are frequently observed in the near-earth plasmas
and in the boundary layers of the earth's magnetosphere \cite{key-0a,key-0b}. A soliton
is a single wave pulse which is generated due to accumulation of
electron density at a particular region. When there is an inhomogeneity
of electrons (or ions) due to the evolution of nonlinear perturbation, it results
in the formation of a potential structure at that specific region,
and henceforth an electric field is generated which is in fact detected
by the spacecraft in the form of a bipolar pulse \cite{key-0b,key-0c,key-0d}. Theoretical 
and observational studies have indicated that these ESWs are basically potential structures
 and weak double layers \cite{key-00}.

In the recent years many physical models have been put forward by
various authors trying to give a correct explanation for the existence
of these solitary structures \cite{key-1a,key-1b,key-1c,key-1d,key-1e,key-4a}.
 In a complex plasmas such as planetary magnetosphere, 
various physical effects play crucial roles in formation of these structures.  
The ambient magnetic field in a plasma
may lead to this pressure anisotropy due to disparate time scales
in parallel and perpendicular directions of the magnetic field when  Coulomb
collisions are sufficiently weak. Many astrophysical plasmas are magnetised and can be considered
almost collisionless where thermal anisotropy is an important factor \cite{key-0e,key-0f}. In near-earth
plasmas, field-aligned electron anisotropies have been found on auroral and magnetospheric plasmas  in both
high and low altitudes \cite{key-0g,key-0h}.

We note that that a full-blown analyses of solitary waves with self-consistent
perturbation of the magnetic field can be very complicated indeed
\cite{key-0i}. 
Various authors have studied the effect of pressure
anisotropy on formation of nonlinear structures in electron-positron-ion
(epi) plasmas \cite{key-2A}, dusty plasmas \cite{key-3A,key-4A,key-5A},
and plasmas with $\kappa$-distributed electrons \cite{key-6A}.
We, in this work, have considered formation of these ion-acoustic
ESWs in the presence of both ion and electron thermal anisotropy,
in the presence of a background magnetic field. In Section 2, we have
presented the plasma model for the formation of these ESWs in a magnetoplasma
with ion pressure anisotropy. In Section 3, we describe a general
procedure for reducing the mathematical equations to study these solitary
structures. In Section 4, we incorporate the electron thermal anisotropy
through a bi-Maxwellian velocity distribution function. In Section
5, we have formulated the pseudo potential analysis and compared our
theoretical results with available observational data. Finally in
Section 6, we summarise our results and conclusions.

\section{The plasma model}

The basic model of our plasma is represented by the following MHD
equations,
\begin{eqnarray}
\frac{\partial n}{\partial t}+\nabla.(n\bm{v}) =  0,\label{eq:continuity}\\
\frac{d\bm{v}}{dt}  = -\frac{e}{m_{i}}\nabla\phi-\frac{1}{m_{i}n}\nabla\cdot{\bm p}+(\bm{v}\times\bm{\Omega}_{i}),\label{eq:momentum}
\end{eqnarray}
which are ion continuity and momentum equations where $n$ is the
ion density, $\Omega_{i}=eB/(m_{i}c)$ is the ion gyro-frequency, and $\bm p$ is
the anisotropic pressure tensor,
\begin{equation}
{\bm p}=p_\perp({\bm I}-{\bm b}{\bm b})+p_\parallel{\bm b}{\bm b},
\end{equation}
with $\bm I$ as the unit dyadic and ${\bm b}={\bm B}/B$ is the unit vector along the field line. The other symbols have their usual meanings. Electrons are assumed to be Boltzmannian. We also invoke the quasi-neutrality
condition 
\begin{equation}
n=n_{e}=F(\phi),\label{eq:fphi}
\end{equation}
where $F$ is a function of plasma potential $\phi$. The ion thermal-anisotropy
is assumed to be described the CGL double adiabatic laws \cite{chew}, 
\begin{eqnarray}
\frac{d}{dt}\left(\frac{p_{\perp}}{\rho B}\right)  =  0,\label{eq:cgl1}\\
\frac{d}{dt}\left(\frac{p_{\parallel}B^{2}}{\rho^{3}}\right)  =  0.\label{eq:cgl2}
\end{eqnarray}
We assume the external magnetic field $\bm{B}$  is in the $\hat{\bm{z}}$
direction and plasma approximation is assumed all throughout. The
application of plasma approximation, in general, implies that the
time scale of perturbation is large enough so that variation of electric
potential $\phi$ due to electrons and ions in space, can be thought
to be smeared out and the scale length over which $\phi$ varies,
is considerably larger than the Debye screening length $\lambda_{D}$.
For our intended parameter regime of magnetosphere of the earth including
auroral regions, this can be justified, where solitary structures
of the order of $\sim10\lambda_{D}$ are observed in the auroral regions
\cite{key-1,key-2,key-3,key-4}.

\subsection{CGL anisotropy}

We note that the double adiabatic equations (CGL theory) \cite{chew}
is rather restrictive in its application as the system \emph{must}
vary sufficiently slowly along the field lines so that particles with
different behaviour at two different points, even along the lines
of force, have only little communication \cite{key-8A}. However,
we should note that the CGL equations are actually a subset of a more
general polybaric pressure equations \cite{key-12A},
\begin{eqnarray}
p_{\perp} & \propto & N^{\gamma}B^{\kappa},\label{eq:polybaric1}\\
p_{\parallel}/p_{\perp} & \propto & N^{\gamma_{a}}B^{\kappa_{a}}.\label{eq:polybaric2}
\end{eqnarray}
The CGL equations can be recovered for $\gamma=\kappa=1$, and $\gamma_{a}=2,\kappa_{a}=-3$.
Measurements from Cluster series of spacecrafts in the earth's magnetosphere
has indicated that the ion anisotropy can be well modelled by Eqs.(\ref{eq:polybaric1},\ref{eq:polybaric2})
\cite{key-9A,key-10A} for different values of the parameters $\gamma,\gamma_{a},\kappa,\kappa_{a}$.
However, our prime objective in this paper has been to incorporate
the ion and electron thermal anisotropies as a \emph{proof of concept}
rather than use it to fully explain observationally obtained experimental
data. We also note that space plasmas including geomagnetic and auroral
plasmas are diverse enough to call for different physical effects
to be incorporated in the theory and in a restricted parameter regime,
the CGL theory is actually found to largely agree with observational
data in solar wind plasmas \cite{key-13A}. 

\section{Reduction of equations}

We now describe a general procedure for reducing Eqs.(\ref{eq:continuity}-\ref{eq:cgl2})
for nonlinear perturbation \cite{key-4a}. We assume an arbitrary electrostatic perturbation
in time and space and define a co-moving co-ordinate $\eta=l_{x}x+l_{z}z-v_{M}t$,
where $l_{x,z}$ are direction co-sines and thus defined by the relation
$l_{x}^{2}+l_{z}^{2}=1$ and $v_{M}$ is the velocity of the nonlinear
wave. Far away from the perturbation, we assume everything to be stationary
and define the boundary conditions  : $\eta\rightarrow\infty$,
$n\rightarrow n_{0}$, $\phi\rightarrow0$, and $v\rightarrow0$.
Without any loss of generality, we can assume that the physical quantities
to be constant along the $\hat{\bm{y}}$ direction. Note that in the
scaled coordinates, we have,
\begin{equation}
\left.\begin{array}{lll}
{\displaystyle \frac{\partial}{\partial t}\equiv\frac{\partial\eta}{\partial t}\frac{\partial}{\partial\eta}}  \equiv  {\displaystyle -v_{M}\frac{\partial}{\partial\eta},}\\
\\
{\displaystyle \frac{\partial}{\partial x}\equiv\frac{\partial\eta}{\partial x}\frac{\partial}{\partial\eta}}  \equiv  {\displaystyle l_{x}\frac{\partial}{\partial\eta},}\\
\\
{\displaystyle \frac{\partial}{\partial z}\equiv\frac{\partial\eta}{\partial z}\frac{\partial}{\partial\eta}}  \equiv 
\displaystyle{l_{z}\frac{\partial}{\partial\eta}.}
\end{array}\right\} \label{eq:coord}
\end{equation}
From the continuity equation we get 
\begin{equation}
l_{x}v_{x}+l_{z}v_{z}-v_{M}=-v_{M}\frac{n_{0}}{n},
\end{equation}
From the $x,y$, and $z$ components of the momentum equation, we get,
\begin{eqnarray}
-v_{M}\frac{n_{0}}{n}v_{x}' = -l_{x}\left[g(n)+h_{\perp}(n)\right]+v_{y}\Omega_{i},\\
-v_{M}\frac{n_{0}}{n}v_{y}' = -v_{x}\Omega_{i},\\
-v_{M}\frac{n_{0}}{n}v_{z}' = -l_{z}\left[g(n)+h_{\parallel}(n)\right],
\end{eqnarray}
where 
\begin{eqnarray}
g(n) = \frac{e}{m_{i}}\phi',\label{eq:gn}\\
h_{\perp}(n) = \frac{p_{\perp0}}{m_{i}nn_{0}}n',\label{eq:fperp}\\
h_{\parallel}(n) = \frac{3p_{\parallel0}n}{m_{i}n_{0}^{3}}n',\label{eq:fpar}
\end{eqnarray}
and $(')$ denotes derivative with respect to $\eta$.

At this point, we would like to introduce the normalisation that we
are going to use. The ion density $n$ is normalised to its equilibrium
density $n_{0}$, $p_{\parallel,\perp}$ to $n_{0}T_{\parallel0,\perp0}$,
velocities to the ion-sound velocity $v_{s}=\sqrt{T_{e0}/m_{i}}$
, potential to $T_{e0}/{e}$, length to the ratio $v_{s}/\Omega$,
and time to $\Omega^{-1}$. The ratio of the ion temperature to the
electron temperature is denoted by $\sigma=T_{0}/T_{e0}$. The normalised
equations are then given by,
\begin{eqnarray}
l_{x}v_{x}+l_{z}v_{z}-v_{M} = -\frac{v_{M}}{n},\label{eq:cont1}\\
-\frac{v_{M}}{{n}}v_{x}' = -l_{x}\left[g(n)+h_{\perp}(n)\right]+v_{y},\label{eq:xmom}\\
-\frac{v_{M}}{n}v_{y}' = -v_{x},\label{eq:ymom}\\
-\frac{v_{M}}{n}v_{z}' = -l_{z}\left[g(n)+h_{\parallel}(n)\right].\label{eq:zmom}
\end{eqnarray}
Differentiating Eqs.(\ref{eq:xmom}) and (\ref{eq:zmom}) with respect
to $\eta$ we get, 
\begin{eqnarray}
-\frac{v_{M}}{n}v_{x}^{\prime\prime}+\frac{v_{M}{n}^{\prime}}{n^{2}}v_{x}^{\prime} = \frac{n}{v_{M}}v_{x}-l_{x}\left[g^{\prime}(n)+h_{\perp}^{\prime}(n)\right],\label{eq:mom2}\\
-\frac{v_{M}}{n}v_{z}^{\prime\prime}+\frac{v_{M}{n}^{\prime}}{n^{2}}v_{z}^{\prime} = -l_{z}\left[g^{\prime}(n)+h_{\parallel}^{\prime}(n)\right],\label{eq:mom3}
\end{eqnarray}
where we have substituted the value of $v_{y}^{\prime}$ from Eq.(\ref{eq:ymom}).
By differentiating Eq.(\ref{eq:cont1}), successively with respect
to $\eta$, we get, 
\begin{eqnarray}
l_{x}v_{x}^{\prime}+l_{z}v_{z}^{\prime} = \frac{v_{M}n^{\prime}}{n^{2}},\label{eq:cont2}\\
l_{x}v_{x}^{\prime\prime}+l_{z}v_{z}^{\prime\prime} = -v_{M}\left(\frac{2n'^{2}}{n^{3}}-\frac{n^{\prime\prime}}{n^{2}}\right).\label{eq:cont3}
\end{eqnarray}\newpage
Using Eqs.(\ref{eq:mom2}-\ref{eq:mom4}), we get, 
\begin{eqnarray}
g^{\prime}(n)+\frac{v_{M}^{2}n'^{2}}{n^{4}}+l_{z}^{2}h_{\parallel}^{\prime}(n)+l_{x}^{2}h_{\perp}^{\prime}(n)=
& \frac{-v_{M}+nv_{M}-nl_{z}v_{z}}{v_{M}}\nonumber\\
&-\,\frac{v_{M}^{2}}{n}\left(\frac{2n'^{2}}{n^{3}}-\frac{n^{\prime\prime}}{n^{2}}\right),\label{eq:mom4}
\end{eqnarray}
where we have used the condition $l_{x}^{2}+l_{z}^{2}=1$ and substituted
the value of $v_{x}$ from Eq.(\ref{eq:cont1}). Eq.(\ref{eq:mom3})
can be integrated to have 
\begin{equation}
v_{z}=c_{1}+\frac{l_{z}}{v_{M}}\int n\left[g(n)+h_{\parallel}(n)\right]\, d\eta\label{eq:vz}
\end{equation}
where $c_{1}$ is the constant of integration to be evaluated by imposing
the boundary conditions. So, finally, using Eqs.(\ref{eq:vz}), from
Eq.(\ref{eq:mom4}), we arrive at a single nonlinear second order
differential equation for the system, 
\begin{eqnarray}
1+g^{\prime}(n)=&-\frac{3v_{M}^{2}n'^{2}}{n^{4}}
-l_{z}^{2}h_{\parallel}^{\prime}(n)-l_{x}^{2}h_{\perp}^{\prime}(n)\nonumber\\
&+\,\frac{n}{v_{M}^{2}}\left\{ v_{M}(v_{M}-c_{1}l_{z})
-l_{z}^{2}\int n\left[g(n)+h_{\parallel}(n)\right]\, d\eta\right\} \nonumber\\
&+\,\frac{v_{M}^{2}n^{\prime\prime}}{n^{3}}.\label{eq:mom5}
\end{eqnarray}

With specific electron distribution and together with plasma approximation,
Eq.(\ref{eq:mom5}) can be written in a generic form as,
\begin{equation}
\alpha(n)n''+\beta(n)n'^{2}+\zeta(n)=0,\label{eq:pot1}
\end{equation}
where $\alpha(n),\beta(n)$, and $\zeta(n)$ are arbitrary functions
of $n$. Eq.(\ref{eq:mom5}) can be re-cast as,
\begin{equation}
\lambda(n)\frac{d^{2}G(n)}{d\eta^{2}}+\zeta(n)=0,\label{eq:pot2}
\end{equation}
where $\lambda(n)$ and $G(n)$ are functions of $n$, to be determined.
The Sagdeev potential $V(n)$ can now be written in terms of $G(n)$
as,
\begin{equation}
V(n)=-\int^{n}\frac{\zeta(n)}{\lambda(n)}G^{\prime}(n)\, dn+c_{2},\label{eq:pot3}
\end{equation}
where $c_{2}$ is an integration constant to be determined by imposing
boundary conditions on $V(n)$.

By comparing the coefficients of $n''$ and $n'^{2}$ of Eqs.(\ref{eq:pot1})
and (\ref{eq:pot2}), we get,
\begin{eqnarray}
\alpha = \lambda(n)\frac{dG(n)}{dn}=\lambda(n)G'(n),\\
\beta = \lambda(n)\frac{d^{2}G(n)}{d\eta^{2}}=\lambda(n) G''(n),
\end{eqnarray}
from which we can write Eq.(\ref{eq:pot1}) as,
\begin{equation}
\alpha(n)G''(n)-\beta(n)G'(n)=0,\label{eq:glambda}
\end{equation}
which determines $G(n)$ and in turn $\lambda(n)$.

\subsection{Isotropic ion pressure}

We note that for isotropic ion pressure, $p\propto\rho^{\gamma}$,
the nonlinear equation, Eq.(\ref{eq:mom5}) becomes \cite{key-4a},
\begin{eqnarray}
1+g^{\prime}(n)= & -\frac{3v_{M}^{2}n'^{2}}{n^{4}}-h^{\prime}(n)\nonumber \\
 & +\,\frac{n}{v_{M}^{2}}\left\{ v_{M}(v_{M}-c_{1}l_{z})-l_{z}^{2}\int n\left[g(n)+h(n)\right]\, d\eta\right\} \nonumber \\
 & +\,\frac{v_{M}^{2}n^{\prime\prime}}{n^{3}}.\label{eq:mom5-1}
\end{eqnarray}
The effect of ion pressure being anisotropic has a direct consequences
on the limits of the Mach number, as we shall see in Sec.5.

\section{Electron velocity distribution}

In
an weakly collisional or collisionless plasma, momenta of various
species can not be effectively exchanged among the field aligned and
perpendicular directions of the ambient magnetic field. This results
in the anisotropic velocity distributions of the particles, say of
electrons and ions. Besides, in the ion-acoustic time scale, which
is of our interest in this work, electron inertia can very well be
neglected. However, these two characteristics 
can be effectively included in the theory by considering the density
of the lighter species be determined solely by an anisotropic velocity distribution,
without considering the momentum balance. This situation is particularly
true in case of equatorially trapped electrons in the geomagnetic
mirror \cite{key-7A}. The anisotropic electron distribution in the
geomagnetic field are experimentally detected by space born experiments
as early as 1979 through the SCATHA and DE-1 spacecraft observations
\cite{key-12}. Electron thermal anisotropy is also observed in the
auroral plasmas \cite{key-14}.

The particle density $n$ for any arbitrary velocity distribution
function $f(\bm{v},\phi)$ in presence of an electrostatic potential
$\phi$, can be obtained from the basic principle,
\begin{equation}
n=\int\!\! f(\bm{v},\phi)\, d^{3}v,
\end{equation}
where the integral is over the entire velocity space. For an anisotropic
distribution function in presence of a magnetic field, which we are
going to consider in this work, the above relation can be written
as,
\begin{equation}
n=\int\!\!\!\int\!\! f\left(v_{\perp}^{2},v_{\parallel},\phi\right)\, dv_{\perp}^{2}dv_{\parallel}.
\end{equation}
We can now apply Liouvilles' theorem to find out the density of particles
at any point along the magnetic field line, which basically states
that the velocity distribution function in a collisionless plasma
is constant along the particle trajectories i.e.\ along the magnetic
lines of force \cite{key-4b,key-4c}. We assume that the total energy ${\cal E}$ and the
adiabatic invariant $\mu$, the magnetic moment are constant throughout
the particle trajectory and so, the distribution function can be entirely
expressed in terms of these variables,
\begin{equation}
n_{s}=\frac{\pi B_{s}}{2}\left(\frac{2}{m}\right)^{3/2}\int\!\!\int\!\!\frac{f({\cal E}-q\phi,\mu)}{({\cal E}-q\phi_{s}-\mu B_{s})^{1/2}}\, d{\cal E}\, d\mu,
\end{equation}
where
\begin{eqnarray}
{\cal E} = \frac{1}{2}m\left(v_{\perp}^{2}+v_{\parallel}^{2}\right)+q\phi,\\
\mu = \frac{1}{2}m\frac{v_{\perp}^{2}}{B},
\end{eqnarray}
and the subscript `$s$' is the field-line label. However, by using
the variables $({\cal E},\mu)$, we have lost the distinction for
the oppositely moving particles along a particular field line, which
can be explicitly taken care of by using two distribution functions
$f^{\pm}$ for particle moving along the line and opposite to it,
$ds/dt\gtrless0$ \cite{key-4b},
\begin{equation}
n_{s}=\frac{\pi B_{s}}{2}\left(\frac{2}{m}\right)^{3/2}\int_{\mu=0}^{\infty}\int_{{\cal E}=q\phi_{s}+\mu B_{s}}^{\infty}\frac{f^{+}+f^{-}}{({\cal E}-q\phi_{s}-\mu B_{s})^{1/2}}\, d{\cal E}\, d\mu.
\end{equation}
Assuming symmetry between the oppositely moving particles, which is
especially true for trapped particles in the equatorial region of
the geomagnetic sphere, we can set $f^{+}=f^{-}=f$,
\begin{equation}
n_{s}=\pi B_{s}\left(\frac{2}{m}\right)^{3/2}\int_{\mu=0}^{\infty}\int_{{\cal E}=q\phi_{s}+\mu B_{s}}^{\infty}\frac{f({\cal E}-q\phi,\mu)}{({\cal E}-q\phi_{s}-\mu B_{s})^{1/2}}\, d{\cal E}\, d\mu.
\end{equation}
If we now take a position  on a field line as a
reference point, we can set $\phi=0$ at that point and $n=n_{0}$,
we can substitute $f({\cal E}-q\phi,\mu)=f({\cal E},\mu)$ in the
above expression to obtain the density at any point along the field
line with reference to the equatorial position,
\begin{equation}
n_{s}=\pi B_{s}\left(\frac{2}{m}\right)^{3/2}\int_{\mu=0}^{\infty}\int_{{\cal E}=q\phi_{s}+\mu B_{s}}^{\infty}\frac{f({\cal E},\mu)}{({\cal E}-q\phi_{s}-\mu B_{s})^{1/2}}\, d{\cal E}\, d\mu.\label{eq:ns}
\end{equation}
Note that the reference position can be set to any convenient position.

However, a clarification, regarding the above expression of density
$n_{s}$ in terms of the magnetic field strength $B_{s}$, must be
made. In principle, the value of $B_{s}$ at a point `$s$' on the
field line with respect to the `$0$' position should be dictated
by experimental observations, say for example, in case of the geomagnetic
field. However, when we would like to obtain the expression for electric
potential $\phi_{s}$ at that position in terms of the density $n_{s}$,
mathematically we would like to express the field strength $B_{s}\equiv B_{s}(\phi)$.

\subsection{Bi-Maxwellian electron distribution}

So far, we have expressed the dependence of electron density $n_{e}$
on plasma potential $\phi$ with an generalised expression Eq.(\ref{eq:fphi}).
We now assume that the electron population can be described by an
anisotropic Maxwellian distribution \cite{key-4d},
\begin{equation}
f\left(v_{\perp}^{2},v_{\parallel}\right)=n_{0}\left(\frac{m_{e}}{2\pi T_{\perp}}\right)\left(\frac{m_{e}}{2\pi T_{\parallel}}\right)^{1/2}\exp\left[-\frac{m_{e}}{2}\left(\frac{v_{\perp}^{2}}{T_{\perp}}+\frac{v_{\parallel}^{2}}{T_{\parallel}}\right)\right].
\end{equation}
As mentioned above, we set $\phi=0$ at the reference point where
$B=B_{0}$ and write the distribution as,
\begin{equation}
f({\cal E},\mu)=n_{0}\left(\frac{m_{e}}{2\pi T_{\perp}}\right)\left(\frac{m_{e}}{2\pi T_{\parallel}}\right)^{1/2}\exp\left[-\frac{\mu B_{0}}{T_{\perp}}-\frac{{\cal E}-\mu B_{0}}{T_{\parallel}}\right].\label{eq:bimax}
\end{equation}
The normalised electron density $n_{e}$ on an arbitrary position
`$s$' on a field line with respect to that in the reference point,
can now be written from Eq.(\ref{eq:ns}) as,
\begin{equation}
n_{e}\equiv F\left(\phi\right)=\gamma(\phi) e^{\phi},\label{eq:ne-bk}
\end{equation}
where we have written $\phi\equiv\phi_{s}$ and substituted the electronic
charge $q=-e$. The factor $\gamma(\phi)$ is given by,
\begin{equation}
\gamma(\phi)=\left[\frac{T_{e\perp}}{T_{e\parallel}}+\left(1-\frac{T_{e\perp}}{T_{e\parallel}}\right)\frac{B_{0}}{B_{s}(\phi)}\right]^{-1}.\label{eq:gamma}
\end{equation}

We note that Eq.(\ref{eq:bimax}) is a generalised case of the anisotropic
bi-$\kappa$ (or bi-Lorentzian) distribution,
\begin{eqnarray}
f({\cal E},\mu) & = & \frac{n_{0}}{\pi^{3/2}\theta_{\perp}^{2}\theta_{\parallel}}\,\frac{\Gamma(\kappa+1)}{\Gamma(\kappa-1/2)}\,\left\{ 1+\frac{2({\cal E}-\mu B_{0})}{m_{e}\kappa\theta_{\parallel}^{2}}+\frac{2\mu B_{0}}{m_{e}\kappa\theta_{\perp}^{2}}\right\} ^{-1-\kappa},\\
\theta_{\parallel,\perp}^{2} & = & \frac{2T_{e\parallel,\perp}}{m_{e}}\,\left(\frac{2\kappa-3}{\kappa}\right)
\end{eqnarray}
in the limit $\kappa\to\infty$. Various experimental observations
indicate that geomagnetic plasmas can be fitted well with the $\kappa$
distribution rather than Maxwellian. This is essentially true for
a beam-plasma system where thermal equilibration time scale for the
particles is less enough so that a full relaxation to a Maxwellian distribution
can not occur during the dynamical time scale. However, auroral electrons
can be well fitted with a bi-Maxwellian distribution \cite{key-15A,key-17A}.
In case of solar wind plasmas also, which is basically a beam-plasma
system, the electron population is found have two distinct distributions
--- a \emph{core} population, which is very well modelled by a bi-Maxwellian
distribution and a \emph{halo} population with super-thermal particles
described by $\kappa$ distribution \cite{key-16A}. In view of this,
we have chosen to consider a bi-Maxwellian electron population for
our work in general, as we expect our results to be relevant in the
auroral plasma regime. This assumption also simplifies our model without
sacrificing the essential details.

\section{The pseudo potential}

We now assume that the quasi-neutrality condition Eq.(\ref{eq:ne-bk})
can be inverted for $\phi$,
\begin{equation}
\phi=F^{-1}(n)
\end{equation}
and can proceed to find out the equivalent Sagdeev potential, following
the formalism outlined in Sec.2.

The density function $g(n)$ in terms of the inverse function can
be written as,
\begin{equation}
g(n)=n'\frac{\partial}{\partial n}F^{-1}(n).
\end{equation}
Following the analysis (see Sec.2), the expression for the Sagdeev
potential can be written as,
\begin{eqnarray}
V(n) = &\int^{n}\left[1+\frac{n}{v_{M}^{2}}\left\{ l_{z}^{2}\left(I_{n}+\frac{3}{4}(n^{4}-1)\sigma_{\parallel}\right)-v_{M}^{2}-l_{z}^{2}I_{1}\right\} \right]\nonumber \\
 &  \times\,\left(3n^{2}l_{z}^{2}\sigma_{\parallel}+\frac{l_{x}^{2}}{n}\sigma_{\perp}+\frac{\partial}{\partial n}F^{-1}(n)-\frac{v_{M}^{2}}{n^{3}}\right)\, dn+c_{2},
\end{eqnarray}
where $c_{2}$ is the integration constant, to be found from the boundary
conditions on $V(n)$. The $I_{n}$ are integrals defined as,
\begin{equation}
I_{n}=\int^{n}n\frac{\partial}{\partial n}F^{-1}(n)\, dn.
\end{equation}

\begin{figure}
\begin{centering}
\includegraphics[width=0.5\textwidth]{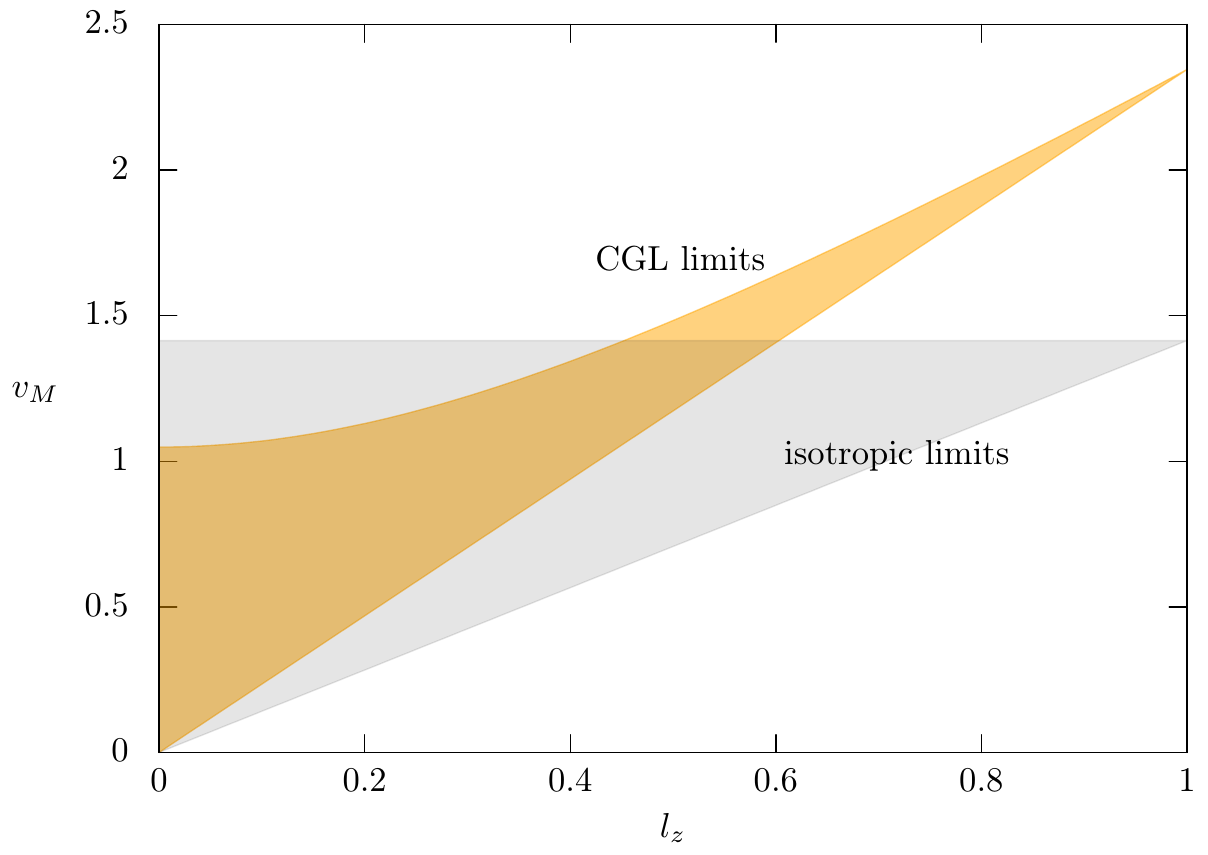} 
\par\end{centering}

\protect\caption{\label{fig:The-limits-of}The limits of Mach number $v_{M}$ for the
CGL thermal anisotropy and isotropic pressure. The anisotropy parameters
are $\sigma_{\perp}=0.1,\sigma_{\parallel}=1.5$ and $\sigma=1$ for
isotropic pressure.}
\end{figure}

\subsection{Limiting Mach number}

The limiting Mach number for formation of solitary structure can be
found out by demanding the condition for local maximum for the pseudo
potential $V(n)$ at $n=1$, which can be conveniently reduced to,
\begin{equation}
\left[l_{z}^{2}\left(\tilde{F}_{1}+3\sigma_{\parallel}\right)-v_{M}^{2}\right]\left[\tilde{F}_{1}+l_{z}^{2}\left(3\sigma_{\parallel}-\sigma_{\perp}\right)+\sigma_{\perp}-v_{M}^{2}\right]<0,
\end{equation}
or
\begin{equation}
l_{z}\sqrt{\tilde{F}_{1}+3\sigma_{\parallel}}<v_{M}<\sqrt{\tilde{F}_{1}+\sigma_{\perp}+l{}_{z}^{2}(3\sigma_{\parallel}-\sigma_{\perp})}.\label{eq:limit}
\end{equation}
where $\tilde{F}_{n}=\partial_{n}F^{-1}(n)$. One can easily see that
no soliton is possible for purely parallel propagation i.e.\ $l_{z}=1$.
For purely perpendicular propagation $(l_{z}=0)$, the condition reduces
to,
\begin{equation}
v_{M}<\sqrt{\tilde{F}_{1}+\sigma_{\perp}}.
\end{equation}
For isotropic ion pressure, condition (\ref{eq:limit}) reduces to
\cite{key-4a},
\begin{equation}
l_{z}\sqrt{\tilde{F}_{1}+\sigma}<v_{M}<\sqrt{\tilde{F}_{1}+\sigma}.\label{eq:limit-1}
\end{equation}
Naturally, this can severely alter the energy regime where a solitary
structure can form as the Mach number actually indicates the total
energy being pumped into a solitary structure. In Fig.\ref{fig:The-limits-of},
we show these two limits for the Mach number.

\begin{figure}
\begin{centering}
\includegraphics[width=1\textwidth]{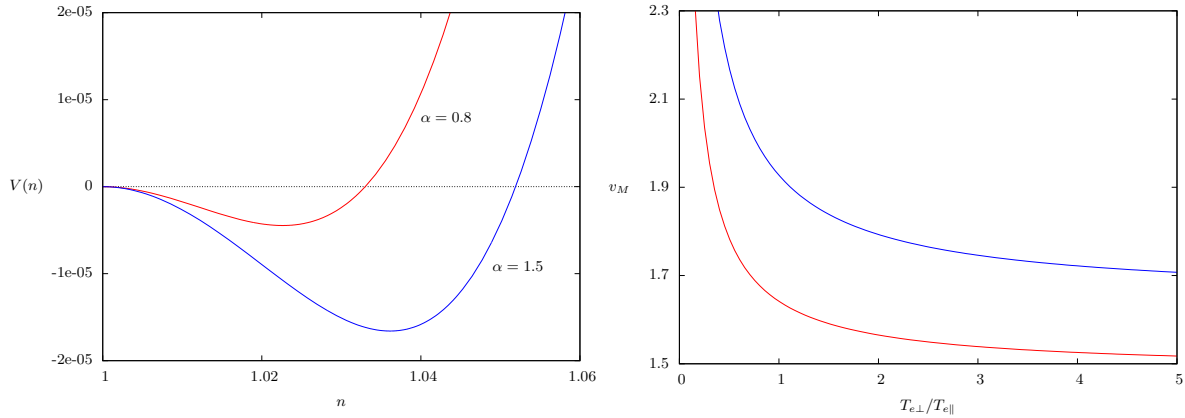}
\par\end{centering}

\protect\caption{\label{fig:The-dependence-of-1}The dependence of pseudo potential
structure (left) and limiting Mach number (right) on the electron
thermal anisotropy parameter $\alpha$. Note that $\alpha>1$ signifies
negative anisotropy ($T_{e\perp}>T_{e\parallel})$. While for large
negative anisotropy the limiting Mach number reaches a constant value,
for large positive anisotropy, the Mach number may reach high values. The ion anisotropy
parameters are $\sigma_\parallel=1.5,\sigma_\perp=1.0$ \cite{key-14}.}
\end{figure}

\subsection{Effect of electron thermal anisotropy}

We note that the effect of electron thermal anisotropy is related
to the magnetic field line variation within the soliton through the
anisotropy factor $\gamma(\phi)$, which is parameterised by the field line
ratio $B_{0}/B_{s}(\phi)$. This justification for this demands an explanation.

The plasma approximation we have used in this work essentially means
that the Debye shielding length is at least a few orders of magnitude
smaller than the size of the nonlinear structures, which is  found
to be correct for the parameter regime of magnetospheric plasmas. This
also ignores the small scale variation of the ambient magnetic field
within the soliton width. In order to model the magnetic field variation
within the soliton width, we consider a state \emph{far away} from
a relaxed plasma state i.e.\ the state of Taylor relaxation \cite{key-4e}, so that
in the moving frame of the soliton, we can write \cite{key-6},
\begin{equation}
\bm{E}\simeq\eta\bm{j},
\end{equation}
which helps us in estimating the electrostatic potential $\phi$,
\begin{equation}
\phi\sim-\eta\int j_{\parallel}\, dl\sim-\eta\frac{BL}{\mu_{0}l}+{\rm const.},\label{eq:bphi}
\end{equation}
where the plasma resistivity can be thought to be a result of field
line stochasticity within the solitary structure. The scale length $l$ is the width of the electric
current structure and $L$ is the length over which the integration
is carried out.

\begin{figure}
\begin{centering}
\includegraphics[width=1\textwidth]{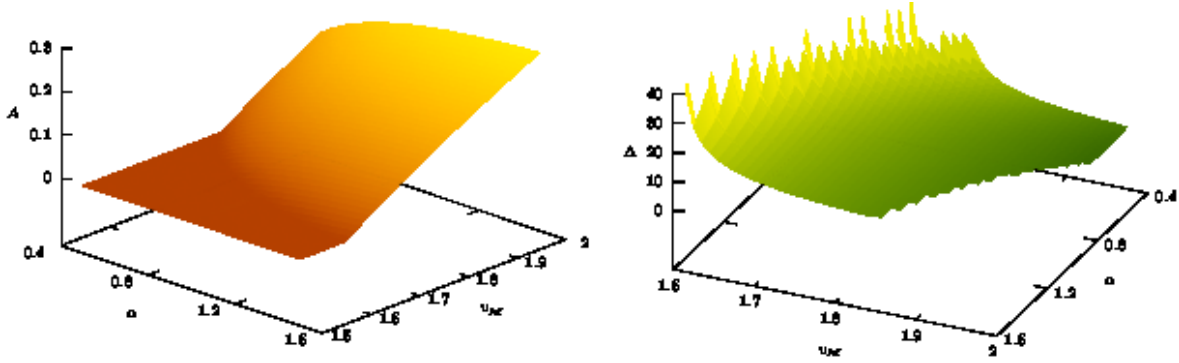}
\par\end{centering}

\protect\caption{\label{fig:The-dependence-of}The dependence of soliton amplitude
$A$ (left) and width $\Delta$ (right) with the electron temperature
anisotropy and Mach number.}
\end{figure}

\begin{figure}
\begin{centering}
\includegraphics[width=1\textwidth]{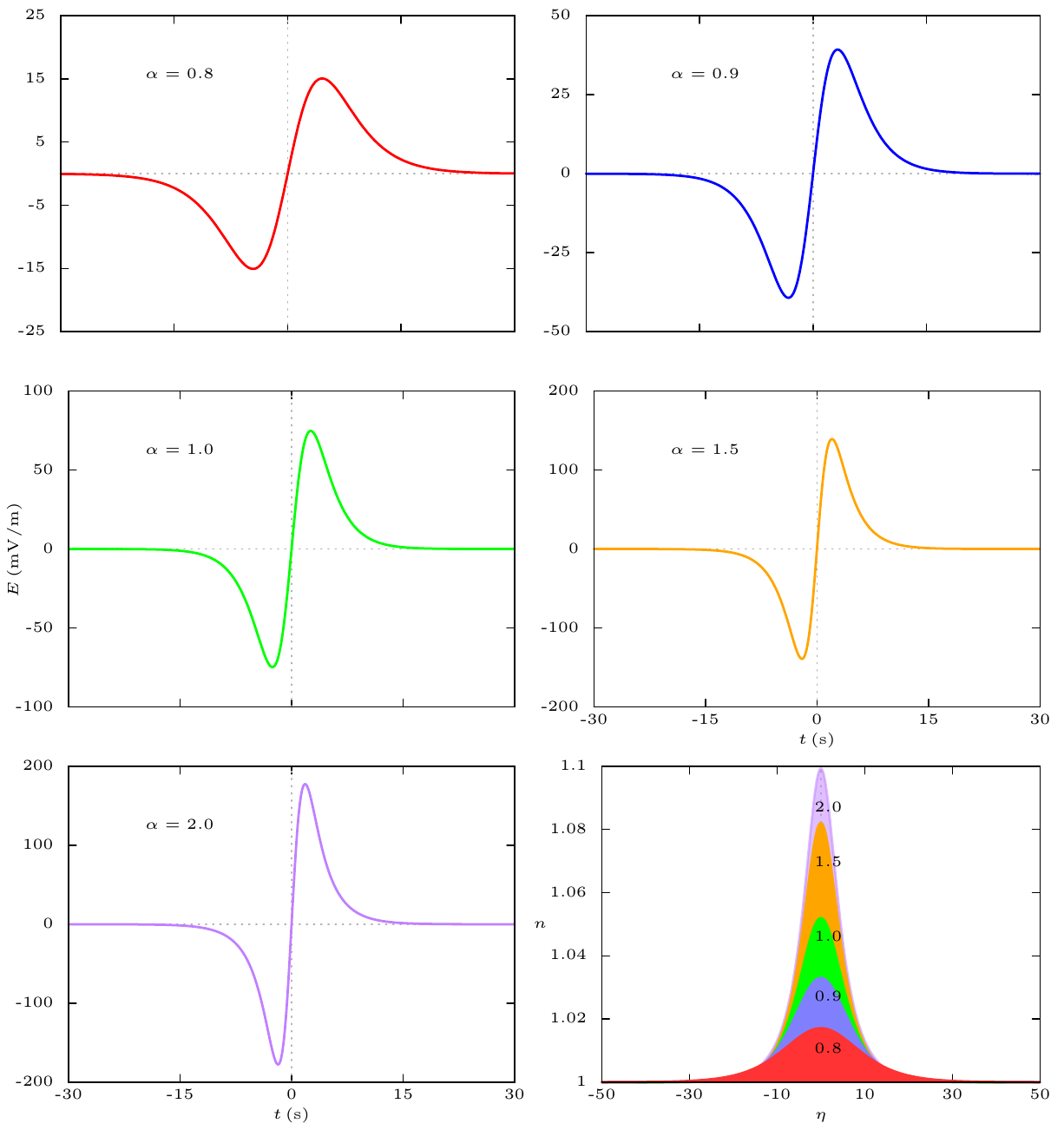}
\par\end{centering}

\protect\caption{\label{fig:Bipolar-electric-field}Bipolar electric field structures,
in the rest frame of the detector, show as they vary in size with
respect to the electron thermal anisotropy parameter $\alpha$. The
bottom right panel shows the corresponding density solitons.
The embedded numbers represent corresponding values of $\alpha$.  The ion anisotropy
parameters are $\sigma_\parallel=1.5,\sigma_\perp=1.0$ \cite{key-14}.}
\end{figure}

We now note that the electron density is given by Eqs.(\ref{eq:ne-bk},\ref{eq:gamma}).
Though we have chosen to parameterise the dependence of magnetic field
line $B_{s}$ on the plasma potential as per the relation (\ref{eq:bphi}),
we note that the resultant equation for $n_{e}$ i.e.\ Eq.(\ref{eq:ne-bk})
becomes analytically non-invertible in terms of $\phi$. So, we approximate
Eq.(\ref{eq:bphi}) for small $\phi$ as,
\begin{equation}
\frac{B_{s}}{B_{0}}\sim e^{-\phi}.
\end{equation}
which also reduces to the correct asymptotic value at the limit $\phi\to0$
in conformation with relation (\ref{eq:bphi}). The plasma potential
is now given by,
\begin{equation}
\phi=\ln\left(\frac{n\alpha}{1-n+n\alpha}\right),\label{eq:plasmapot}
\end{equation}
where $\alpha=T_{e\perp}/T_{e\parallel}$ is a measure of electron
temperature anisotropy. With this, the bounds on limiting Mach number
becomes,
\begin{equation}
l_{z}\left(\frac{1}{\alpha}+3\sigma_{\parallel}\right)^{1/2}<v_{M}<\left[\frac{1}{\alpha}+l_{z}^{2}(3\sigma_{\parallel}-\sigma_{\perp})+\sigma_{\perp}\right]^{1/2}
\end{equation}
which becomes
\begin{equation}
\left.\begin{array}{rclr}
l_{z}\left(3\sigma_{\parallel}\right)^{1/2}\,< & v_{M} & <\,\left[l_{z}^{2}(3\sigma_{\parallel}-\sigma_{\perp})+\sigma_{\perp}\right]^{1/2}, & \qquad\alpha\gg1,\\
\\
l_{z}\alpha^{-1/2}\,< & v_{M} & <\,\alpha^{-1/2}, & \qquad\alpha\ll1.
\end{array}\right\} 
\end{equation}
A pseudo potential formation and the dependence of the limiting Mach
number on the thermal anisotropy parameter $\alpha$, are shown in
Fig.\ref{fig:The-dependence-of-1}. As we can see that for large $\alpha$
(large negative anisotropy), the allowed Mach number interval becomes
constant, whereas for small $\alpha$ (large positive anisotropy),
the interval becomes progressively narrower with Mach number reaching
a very high value. If we analyse the soliton amplitude and width,
we observe that soliton amplitude $A$ increases with $\alpha$ and
Mach number and the width decreases with Mach number while it is largely
independent on $\alpha$. Physically we expect to see large amplitude
and slow moving soliton in the region of large negative anisotropy.
The dependence of soliton amplitude $A$ and width $\Delta$ with
the anisotropy parameter and Mach number are shown in Fig.\ref{fig:The-dependence-of}.

Note that the soliton amplitude $A$ is given by first zero of the pseudo
potential away from $n=1$. The width $\Delta$ is related to the
maximum depth $d$ of the soliton as $\Delta=A/\sqrt{d}$. The shallower
is the potential, the narrower is the soliton.

\section{Comparison with experimental data}

In this section, we try to interpret certain experimental results.
However, our results should only indicate an overall trend in the
observational data rather than definitively reproducing experimental
results. 
A major contribution to our knowledge about temporally and spatially
localised electrostatic structures in the auroral regions are due
to the Freja and FAST satellite missions in the 90's. These satellite
missions clearly demonstrated the existence of such structures \cite{key-7}.
More recently, observations by Cluster spacecrafts at a geocentric
distance of about $\sim5R_{E}$ gave a good idea about detailed parameters
of these structures \cite{key-8}. Some typical parameters of these
localised structures are peak-to-peak electric field variations of
about $\sim30-170\,{\rm mV}/{\rm m}$ with a life time ranging from
$10-280\,{\rm secs}$ \cite{key-8}. 
Note that the nonlinear structures observed in these near-earth plasmas
actually fall into the ion-acoustic time scale which also includes
the Alfv\'en time scale and this has been confirmed by various experimental
data as well as theoretical works \cite{key-14A}. 

On the other hand there are strong evidences of electron thermal anisotropy
in the auroral regions, which are due to several satellite based investigations
viz.\ IMP 6 \cite{key-9}, AMPTE/CCE \cite{key-10}, AMPTE/IRM, \cite{key-11}
and SCATHA \cite{key-12}. Typically, electron thermal anisotropy
is present almost all throughout across the range of electron energies
$\sim0.1-10\,{\rm keV}$. In most cases, a positive anisotropy $(T_{e\parallel}>T_{e\perp})$
is observed in lower energies up to $\sim1\,{\rm keV}$ and negative
anisotropy $(T_{e\parallel}<T_{e\perp})$ is observed toward the higher
electron energy $\sim10\,{\rm keV}$ \cite{key-14}. In order to compare
these results, we transform the density soliton to a bipolar electric
field structure, which is what detected on board these satellites.
From the  relation (\ref{eq:plasmapot}), we can find out the
equivalent electric field as,
\begin{equation}
E=-\frac{d\phi}{d\eta}=-\frac{1}{[n+(\alpha-1)n^2]}\frac{dn}{d\eta}.
\end{equation}
Note that for the electron anisotropy parameter $\alpha=1$ (isotropic), the relation reduces to the classical
Boltzmannian relation. The width of these structures can be transformed to rest frame of
the satellite in terms of pulse duration, as reported experimentally.
With the normalisation we have used, we now plot these bipolar structures
for various values of electron thermal anisotropy along with the soliton
structures in Fig.\ref{fig:Bipolar-electric-field}. In our calculations,
 we have assumed an average electron temperature of $\sim10\,{\rm keV}$,
which is typical for these parameters of negative anisotropy $(\alpha>1.0)$
in these regions \cite{key-14}. The Mach number is fixed at $1.7$
for which we get the width of these structure in the rest frame of
the detector in the order of $\sim30\,{\rm s}$ (in terms of pulse
life-time).
These numbers largely agree with observed data in the auroral regions
as reported by these satellite based observations. As we can see from  
Fig.\ref{fig:Bipolar-electric-field} that as the electron anisotropy parameter increases, the 
peak-to-peak variation of the electric field becomes more, we can conclude that
the electron thermal anisotropy has a decisive role in determining the amplitude
of these electrostatic structure. However, we still do not have  direct observational data about these
electrostatic structures in auroral plasmas measured in relation to the electron thermal anisotropy.

\section{Summary and conclusions}
In this work, we have analysed the formation of electrostatic solitary structures through a pseudo potential
analysis in a magnetoplasma with both electron and ion anisotropy which is induced by the ambient  magnetic field along with a weak collisional regime. The ion thermal anisotropy is modelled through a CGL double adiabatic equations of state while the electron anisotropy is incorporated with a bi-Maxwellian distribution. In the ion-acoustic time scale, the electrons are considered inertia-less. We have shown that a self-consistent
inclusion of both ion and electron thermal anisotropies can possibly explain large amplitude electrostatic structures observed in the region of negative electron thermal anisotropy in the magnetosphere.

However, we note that the CGL double adiabatic theory has only very limited application in a restricted parameter regime. Nevertheless, it does show us in principle, the effect of ion thermal anisotropy. It also calls for a more comprehensive analysis which involves realistic pressure anisotropy such as polybaric equations of state
\cite{key-12A}.

\section*{Acknowledgement}
One of the authors, MK acknowledges UGC for fellowship grant (RFSMS). The authors would like to thank two anonymous referees for making the work more informative.


\section*{References}
\begin{thebibliography}{99}
\bibitem{key-0a}Moola S, Bharuthram R, Singh S V and Lakhina G S
2003 \textit{Pramana} \textbf{61} 1209

\bibitem{key-0b}Matsumoto H, Kojima H, Omura Y, Okada M, Nagano I
and Tsutsui 1994 \textit{Gephys.\ Res.\ Lett.} \textbf{21} 2915

\bibitem{key-0c}Franz J R, Kintner P M and Pichett J S 1998 \textit{Geophys.\ Res.\ Lett.}
\textbf{25} 1277

\bibitem{key-0d}Ergun R E et al.\ 1998 \textit{Geophys.\ Res.\ Lett.}
\textbf{25} 2041

\bibitem{key-00}Picket J S et al.\ 2008 \textit{Adv.\ Space Res.}
\textbf{41} 1666

\bibitem{key-1a}Tsurutani B T, Arballo J K, Lakhina G S, Ho C M,
Buti B, Pickett J S and Gurnett D A 1998 \textit{Geophys.\ Res.\ Lett.}
\textbf{25} 4117

\bibitem{key-1b} Lakhina G S, Tsurutani B T, Kojima H and Matsumoto
H 2000 \textit{J.\ Geophys.\ Res.} \textbf{105} 27791

\bibitem{key-1c}Franz J R, Kintner P M, Pickett J S and Chen L-J
2005 \textit{J.\ Geophys.\ Res.} \textbf{110} A09211

\bibitem{key-1d}Williams J D, Chen L-J, Kurth W S, Gurnett D A, Dougherty
M K and Rymer A M 2005 \textit{Geophys.\ Res.\ Lett.} \textbf{32}
L17103

\bibitem{key-1e}Bora M P, Choudhury B and Das G C 2012 \textit{Astrophys.\ Space
Sci.} \textbf{341} 515

\bibitem{key-4a}Choudhury B, Goswami R, Das G C and Bora M P 2013
\textit{Phys.\ Plasmas} \textbf{20} 042902

\bibitem{key-0e}Blanc M, Kallenbach R and Erkaev N V 2005 \textit{Space
Sci.\ Rev.} \textbf{116} 227

\bibitem{key-0f}Schekochihin A A, Cowley S C, Kulsrud R M and Sharma
P 2005 \textit{ApJ} \textbf{629} 139

\bibitem{key-1s}Seough J, Yoon P H, Kim K-H and Lee D-H 2013 \emph{Phys.\ 
Rev.\ Lett.}\textbf{110} 071103

\bibitem{key-0g}Sharp R D, Shelley E G, Johnson R G and Ghielmetti
A G 1980 \textit{J.\ Geophys.\ Res.} \textbf{85} 92

\bibitem{key-0h}Collin H L, Sharp R D and Shelley E G 1982 \textit{J.\ Geophys.\ Res.}
\textbf{87} 7504

\bibitem{key-0i}Meuris P and Verheest F 1996 \textit{Phys.\ Lett.\ A}
\textbf{219} 2992

\bibitem{key-2A}Adnan M, Williams G, Qamar A, Mahmood S and Kourakis
I 2014 \emph{Eur.\ Phys.\  J.\ D }\textbf{68} 247

\bibitem{key-3A}Choi C R, Ryu C-M, Lee D-Y, Lee N C and Kim Y-H 2007
\emph{Phys.\ Lett.\ A} \textbf{364} 297

\bibitem{key-4A}Choi C R, Ryu C-M, Lee N C, Lee D-Y and Kim Y-H 2005
\emph{Phys.\ Plasmas }\textbf{12} 07231

\bibitem{key-5A}Adnan M, Mahmood S and Qamar A 2014 \emph{Contrib.\ 
Plasma Phys.}\textbf{54} 724

\bibitem{key-6A}Adnan M, Mahmood S and Qamar A 2014 \emph{Adv.\  Space
Res.}\textbf{53} 845

\bibitem{key-6AA}Chuang S-H, Hau L-N 2009 \emph{Phys.\  Plasmas} \textbf{16} 022901

\bibitem{chew}Chew G F, Goldberger M L and Low F E 112 \textit{Proc.\ R.\ Soc.\ A}
\textbf{236} 112

\bibitem{key-1}Carlson C W, McFadden J P, Ergun R E et al.\ 1998
\textit{Geophys.\ Res.\ Lett.} \textbf{25} 2017

\bibitem{key-2}McFadden J P et al.\ 1998 \textit{Geophys.\ Res.\ Lett.}
\textbf{25} 2025

\bibitem{key-3}Ergun E R, Su Y J, Anderson L et al.\ \textit{Phys.\ Rev.
Lett.} \textbf{87} 045003

\bibitem{key-4}Anderson L, Ergun R E, Main D et al.\ 2002 \textit{Phys.\ Plasmas}
\textbf{9} 3600

\bibitem{key-8A}Kulsrud R M 1982 \emph{Handbook of Plasma Physics
}\emph{
(Eds M N Rosenbluth and R Z Sagdeev), North-Holland 115}

\bibitem{key-12A}Stasiewicz K 2005 \emph{Phys.\ Rev.\ Lett.}\textbf{95}
015004

\bibitem{key-9A}Stasiewicz K 2004 \emph{Gephys.\ Res.\ Lett.}\textbf{31}
L21 804

\bibitem{key-10A}Stasiewicz 2004 \emph{Phys.\ Rev.\ Lett.} \textbf{93}
125004

\bibitem{key-13A}Schulz M and Eviatar A 1973 \emph{J.\ Geophys.\ Res.}\textbf{78} 3948

\bibitem{key-7A}Olsen R C 1981 \emph{J.\ Geophys.\ Res.}\textbf{86
}11235

\bibitem{key-12} Richardson J D, Fennell J F and Croley Jr D R 1981
\textit{J.\ Geophys.\ Res.} \textbf{86} 10105

\bibitem{key-14}Janhunen P, Olsson A, Laakso H and Vaivads A 2004
\textit{Ann.\ Geophys.} \textbf{22} 237

\bibitem{key-4b}Whipple Jr E C 1977 \textit{J.\ Geophys.\ Res.}
\textbf{82} 1525

\bibitem{key-4c}Moncuquet M, Bagenal F and Meyer-Vernet N 2002 \textit{J.\ Geophys.\ Res.}
\textbf{107} SMP 24-1

\bibitem{key-4d}Huang T S and Birmingham T J 1992 \textit{J.\ Geophys.\ Res.}
\textbf{97} 1511

\bibitem{key-15A}Marghitu O, Klecker B and McFadden J P 2006 \emph{Adv.\ 
Space Res.}\textbf{38} 1694

\bibitem{key-17A}Janhunen P and Olsson A 2008 \emph{Ann.\  Geophysicae
}\textbf{16} 292

\bibitem{key-16A}\v{S}tver\'ak \v{S}, Tr\'avin\'i\u{c}ek P, Maksimovic
M, Marsch E, Fazakerley A N and Scime E E 2008 \emph{J.\  Geophys.\  Res.}\textbf{113} A03103

\bibitem{key-4e}Taylor J B 1974 \textit{Phys.\ Rev.\ Lett.} \textbf{33}
1139

\bibitem{key-6} Wilmot-Smith A L, Hornig G and Pontin D I 2009 \textit{ApJ}
\textbf{696} 1339

\bibitem{key-7}Marklund G, Blomberg L, Fälthammar C-G, Lindqvist
P-A and Eliasson L 1995 \textit{Ann.\ Geophysicae} \textbf{13} 704

\bibitem{key-8}Marklund G T et al.\ 2004 \textit{Nonlin.\ Processes
Geophys.} \textbf{11} 709

\bibitem{key-14A}Ekeberg J, Wannberg G, Eliasson L and Stasiewicz
K 2010 \emph{Ann.\ Geophys.}\textbf{28} 1299

\bibitem{key-9}Hada T, Nishida A, Terasawa T and Hones Jr E W 1981
\textit{J.\ Geophys.\ Res.} \textbf{86} 11211

\bibitem{key-10}Klumpar D M, Quinn J M and Shelley E G 1988 \textit{J.\
Geophys.\ Res.} \textbf{93} 1295

\bibitem{key-11}Sergeev V A, Baumjohann W and Shiokawa K 2001 \textit{Geophys.\
Res.\ Lett.} \textbf{28} 3813
\end{thebibliography}
\end{document}